\documentclass[a4paper]{article}

\usepackage{INTERSPEECH2022}

\usepackage{microtype}
\usepackage{graphicx}
\usepackage{subfigure}
\usepackage{booktabs} 

\usepackage{hyperref}       
\usepackage{url}            
\usepackage{amsmath}
\usepackage{amssymb}
\usepackage{mathtools}
\usepackage{amsthm}
\usepackage[makeroom]{cancel}

\usepackage{amsfonts}       
\usepackage{nicefrac}       
\def\mcal{\mathcal}
\usepackage{xcolor}         

\DeclarePairedDelimiterX{\norm}[1]{\lVert}{\rVert}{#1}

\title{Boosting Self-Supervised Embeddings for Speech Enhancement}
\name{Kuo-Hsuan Hung$^1$, Szu-wei Fu$^2$, Huan-Hsin Tseng$^3$, Hsin-Tien Chiang$^3$, Yu Tsao$^3$, Chii-Wann Lin$^1$}
\address{
  $^1$National Taiwan University, $^2$Microsoft Corporation, $^3$Academia Sinica
}
\email{\{d07528023, d04922007\}@ntu.edu.tw\\ \{htseng, coffee091524, yu.tsao\}@citi.sinica.edu.tw\\
cwlinx@ntu.edu.tw}

\begin{document}

\maketitle{}

\begin{abstract}
 Self-supervised learning (SSL) representation for speech has achieved state-of-the-art (SOTA) performance on several downstream tasks. However, there remains room for improvement in speech enhancement (SE) tasks. In this study, we used a cross-domain feature to solve the problem that SSL embeddings may lack fine-grained information to regenerate speech signals. By integrating the SSL representation and spectrogram, the result can be significantly boosted. We further study the relationship between the noise robustness of SSL representation via clean-noisy distance (CN distance) and the layer importance for SE. Consequently, we found that SSL representations with lower noise robustness are more important. Furthermore, our experiments on the VCTK-DEMAND dataset demonstrated that fine-tuning an SSL representation with an SE model can outperform the SOTA SSL-based SE methods in PESQ, CSIG and COVL without invoking complicated network architectures. In later experiments, the CN distance in SSL embeddings was observed to increase after fine-tuning. These results verify our expectations and may help design SE-related SSL training in the future.
 
\end{abstract}


\section{Introduction}
Speech is an effective and efficient way of communication between individuals, playing an essential role in human-computer interactions. However, during communication, there is often undesired interference from the environment and surroundings, such as \textit{environmental noise}, \textit{background noise} and \textit{reverberations}, so that speech quality and intelligibility often degrade. The process of reducing the background noise with optimal preservation of the original speech quality is then referred to as speech enhancement (SE).


With recent developments in deep learning (DL), deep learning-based SE models have mostly outperformed traditional SE methods~\cite{xu2014regression,lu2013speech, wang2018supervised}. 
The DL-based SE framework generally concerns input features~\cite{wang2021speech}, advanced SE models~\cite{valin2020perceptually, choi2018phase,qi2020tensor}, objective functions~\cite{kumar2016speech, sivaraman2021personalized}, and data augmentations~\cite{kim2021specmix}. This study aims to evaluate the relationship and impact of input features regarding DL-based SE performance.

Self-supervised learning (SSL) utilizes a large amount of \textit{unlabeled} data to extract meaningful representations~\cite{liu2022audio}. In many applications, supervised learning generally outperforms unsupervised learning. However, collecting a large amount of labeled data is time-consuming and sometimes unrealistic. Therefore, the SSL can be leveraged in the circumstances with amounts of unlabeled data to provide expressive (latent) representations and use these latent features as inputs for downstream tasks. It has been verified in various fields that the SSL improves the performance of downstream tasks. Particularly, a few promising SSL models have been proposed for speech-related tasks. One major application is \textit{speech recognition}, where the contributing SSL models include contrastive predictive coding (CPC)~\cite{oord2018representation}, wav2vec~\cite{schneider2019wav2vec}, and HuBERT~\cite{hsu2021hubert}. Recently, there have been some application scenarios where the output representation of an SSL model is used to replace the conventional (data) feature and it turns out to achieve better performance than the original input features~\cite{wang2021fine,cooper2022generalization,nguyen2020investigating}. 

Currently, there are only few studies applying SSL features to SE. Huang et al.~\cite{ZiliSSL} proposed applying SSL features to SE, where the authors observed that when training with weighted-sum representations \emph{``for most SSL models, lower layers generally obtain higher weights.''}. This may be because `\emph{`some local signal information necessary for speech reconstruction tasks is lost in the deeper layers''}. In this study, to solve the aforementioned problem, we propose two simple solutions: 1) utilizing \textit{cross-domain features} as model inputs to compensate for the information loss in SSL features, 2) fine-tuning an SSL model together with an SE model such that the extracted SSL features can derive useful information for SE. Additionally, we analyze the noise-robustness property of SSL features and provide some insight into the relationship with SE. In summary, without introducing complicated or advanced models, our results are comparable to those of state-of-the-art (SOTA) SE methods in the VCTK-DEMAND dataset.


\section{Related Work}
In this section, we first briefly review some commonly-used SSL models and studies using cross-domain features and fine-tuned SSLs in other specific tasks. Related research using SSL on speech enhancement is also surveyed.

\subsection{SSL models}
The SSL model can be categorized into generative modeling, discriminative modeling and multi-task learning. \textbf{Generative modelling} extracts the data input to the representation by the encoder and reconstruct it back to the input by the decoder. These include APC~\cite{chung2019unsupervised}, Mockingjay~\cite{liu2020mockingjay}, DecoAR 2.0~\cite{ling2020decoar}, and Audio2Vec~\cite{tagliasacchi2020pre}. 
\textbf{Discriminative modeling} extracts the input into an representation and measures the corresponding similarity. Such studies include CPC, HuBERT~\cite{hsu2021hubert}, WavLM~\cite{chen2021wavlm} and SPIRAL~\cite{huang2022spiral}. One of the most representative works for the \textbf{multi-task learning} approach is PASE+~\cite{ravanelli2020multi}, which picks up a meaningful speech representation capable of the multi-tasking objective.
In this study, we adopted three SSL models to extract the latent representations: \textit{Wav2vec 2.0}, \textit{HuBERT} and \textit{WavLM}, where they all achieved excellent performances in SUPERB~\cite{yang2021superb}, a challenge to gauge the performance of SSL models under different speech tasks. 



\subsection{Cross-domain feature and fine-tuning SSL}\label{subsec: fine-tuning SSL}

Studies \cite{chen2021wavlm} and \cite{zezario2021deep} have shown that the cross-domain feature can help increase task accuracy on ASR and speech assessment metrics.
Additionally, fine-tuning an SSL on the downstream task has also been shown to provide a significant improvement. In wav2vec 2.0, the SSL model was fine-tuned on label data with the CTC loss for downstream recognition tasks. There is other literature fine-tuning SSL models for non-ASR speech tasks, \textit{e.g.,} on speech emotion recognition~\cite{wang2021fine}, spoken language understanding ~\cite{nguyen2020investigating}, and MOS prediction~\cite{cooper2022generalization}.
The research above showed that fine-tuning the SSL or combining SSL embeddings with raw features can achieve better performance.


\subsection{SSL for SE}
Self-supervised pre-trained models have been increasingly applied to many speech-related tasks, including SE~\cite{wang2020self, 9053925}. Several works PFPL~\cite{hsieh2020improving}, PERL~\cite{kataria2021perceptual}, and K-SENet~\cite{sun2021boosting} applied SSL pretrained models as perceptual loss.
Some other works extracted latent representations as SE model input, such as~\cite{ZiliSSL,tsai2022superb} extracted the latent variables and evaluated 11 SSL upstream methods on the SE downstream task. SSPF~\cite{qiu2021self} utilized phonetic characteristics into a
deep complex convolutional network via a CPC model pre-trained with self-supervised
learning.


\section{The Proposed Method}

\subsection{Incorporate spectrogram with SSL embedding as input}
\begin{figure}[t]
  \centering
  \includegraphics[width=\linewidth]{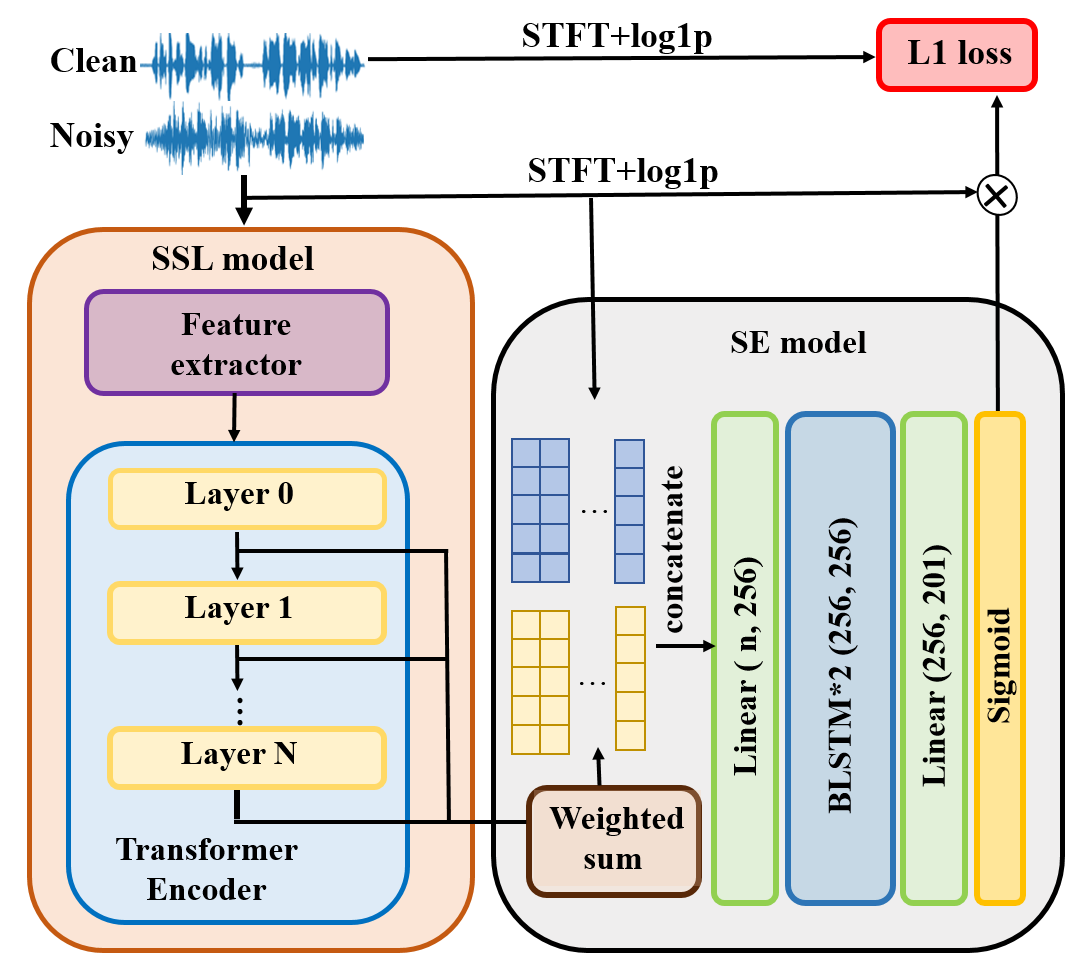}
  \caption{Flowchart of the proposed method with cross-domain feature to boost the performance of SE model.}
  \label{fig:model}
\end{figure}

In~\cite{ZiliSSL}, the authors utilized the SSL representation on the SE as the downstream task. When training with weighted-sum representations, they found that the lower layers generally obtained higher weightings. In additional, \cite{9688093} analyzed the layer-wise representation of Wav2vec 2.0, which showed that the transformer layers in Wav2vec 2.0 followed an autoencoder-style behavior. The latent representation in the lower layers correlates with raw acoustic features such as FBANK. Therefore, combining these two observations, we reasoned that for generation tasks such as SE, the raw acoustic feature should be provided to compensate for the fine-grained deficiencies. In this study, we assess the combination of the $\log 1p$~\cite{fu2020boosting}\footnote{The $\log 1p$ feature is referred to as the scaled transformation $(\log 1p)(X_{tf}):= \log (1 + X_{tf}) \geq 0$ for a spectrogram amplitude $X = (X_{tf})$.} spectrogram feature with SSL representation. The model architecture is illustrated in Fig.~\ref{fig:model}. The noisy waveform is first fed into the SSL model to generate a latent representation, which is subsequently concatenated with the noisy $\log 1p$ feature as the model input. The enhancement prediction is therefore obtained by multiplying the model output with noisy $\log 1p$ features. In the inference stage, a noisy phase is applied to reconstruct the enhanced waveform.

\subsection{Fine-tune SSL model with SE}\label{subsec:Finetune SSL model}
Fine-tuning the pre-trained SSL model with the downstream tasks can generally obtain better results as mentioned in Sec.~\ref{subsec: fine-tuning SSL}. In this study, we follow ~\cite{wang2021fine} to fine-tune the pre-trained SSL model in two ways: partial fine-tuning (PF) and entire fine-tuning (EF). SSL models are generally separated into CNN-based feature extractors and transformer-based encoders. For PF, only the transformer-based encoder is fine-tuned during downstream training. As for EF, both the feature extractor and the encoder are fine-tuned. To our best knowledge, this is the first study applying various fine-tuning SSL models for SE tasks.

\subsection{Noise robustness analysis}\label{subsec:Noise robustness analysis}
Noise robustness is usually desired when training the SSL. Therefore, we intended to investigate whether a representation equipped with noise robustness may help improve SE. Given an SSL model $f$, we define the notion of the distance between the latent variables of noisy speech and clean speech as:
\begin{equation}\label{E:distance}
d_{f, \ell} \left( \mathbf{Z}^{(\ell)}_c,  \mathbf{Z}^{(\ell)}_n \right) = \frac{1}{T} \sum_{t=1}^T \norm{ g^{(\ell)}(\mathbf{z}^{(\ell)}_c(t)) - g^{(\ell)}(\mathbf{z}^{(\ell)}_n(t)) }^2
\end{equation}
where $\ell \in \mathbb{N}$ is the layer depth, $\mathbf{Z}^{(\ell)}_c  := \{ \mathbf{z}^{(\ell)}_c (t) \}_{t=1}^T $ and $\mathbf{Z}^{(\ell)}_n := \{ \mathbf{z}^{(\ell)}_n (t)\}_{t=1}^T $ denote the collection of clean and noisy latent representations of layer $\ell$ in model $f$ and frame $t$, respectively. Particularly, $\ell=0$ corresponds to the output of feature extractor (Fig.~\ref{fig:model}). A normalization function $g^{(\ell)}$ on each layer $\ell$ is deployed to normalize the latent features $\mathbf{z}^{(\ell)}(t)$ for balancing scales and ensuring equal comparisons. Specifically, we shall utilize $g^{(\ell)}(\mathbf{z}^{(\ell)}(t)) := (\mathbf{z}^{(\ell)}(t) - \mu_{\ell}) / \sigma_{\ell}$ with $\mu_{\ell}$ and $\sigma_{\ell}$ denoting the mean and variance of the latent points $\mathbf{Z}^{(\ell)} := \{ \mathbf{z}^{(\ell)}(t) \}_{t=1}^T$. This study computes the layer-wise distance $d_{f, \ell}( \mathbf{Z}^{(\ell)}_c,  \mathbf{Z}^{(\ell)}_n )$ as the \emph{noise robustness}, so that when the distance $d_{f, \ell}$ is small, this layer $\ell$ is regarded as noise robust. In addition to noise robustness, we further consider the layer weighting as an \textit{importance index} for the SSL layers on SE via the weighted-sum approach~\cite{yang2021superb}:
\begin{equation}\label{E:weightedsum}
\mathbf{Z}_{\text{WS}} := \sum_{\ell=0}^{L-1}w(\ell)\, \mathbf{Z}^{(\ell)}
\end{equation}
with parameters $w(\ell) \geq 0 $, $ \sum_{\ell} w(\ell) = 1$ denoting the weight of layer $\ell$ determined by the network and $\mathbf{Z}^{(\ell)}$ is required to keep same dimension for all layers. Later in Sec.~\ref{subsec: Noise robustness}, a bundle of CN distance curves shall be computed to analyze the averaging tendency under stochastic training process with different SSL models. Namely, a SSL model $f$ with randomly sampled inputs $(\mathbf{X}_c, \mathbf{X}_n)$ drawn from the entire training set yield $\ell \mapsto d_{f, \ell} ( \mathbf{Z}^{(\ell)}_c,  \mathbf{Z}^{(\ell)}_n )$ by Eq.~(\ref{E:distance}). Thus, random sampling from data distribution $\mcal{P}$ gives an averaging CN distance curve,
\begin{equation}\label{E:avg CN distance}
\overline{d}_{f, \ell} =  \mathop{\mathbb{E}}_{(\mathbf{X}_c, \mathbf{X}_n) \sim \mathcal{P}} \left[ d_{f, \ell} \left( \mathbf{Z}^{(\ell)}_c,  \mathbf{Z}^{(\ell)}_n \right) \right]
\end{equation}
Experiments will be set up in the next section to verify our boosting method and the relationship with SSL latent representations.

\section{Experiments}

\subsection{Datasets and Evaluation metrics}
A commonly used premixed dataset VCTK-DEMAND~\cite{valentini2016investigating} was adopted to evaluate our method. There were 11,572 utterances with four signal-to-noise ratios (SNRs) (15, 10, 5, and 0 dB) in the training set and 824 utterances with four SNRs (17.5, 12.5, 7.5, and 2.5 dB) in the testing set. The experimental results are assessed with wideband PESQ and STOI for speech quality and intelligibility. Another three widely used metrics: CSIG, CBAK and COVL are applied to measure signal \textit{distortion}, \textit{noise distortion}, and \textit{overall quality evaluation}, respectively. Our implementation is available at \url{https://github.com/khhungg}.

\subsection{Model Structures}
A BLSTM is adopted as an SE model for light weight purpose with decent performance. The model architecture is depicted in Fig.~\ref{fig:model}, consisting of 
(a) 2 linear layers, (b) two-layered BLSTM of 256 hidden units and (c) a sigmoid activation to generate the prediction mask. During the training stage, to obtain fixed-length data within a batch, each utterance was randomly sampled as 20,480 samples (128 frames $\times$ 160 hop length). The signal approximation (SA) method~\cite{liu2019supervised} was used to estimate the target spectrum via the mask prediction. $L_1$-loss and the Adam optimizer were deployed for the SE model, along with a random splitting of training and validation set by $95\%$ and $5\%$ ratio.

\subsection{Results}
\subsubsection{Including spectrogram as extra input features}
In a previous study~\cite{ZiliSSL}, only fixed self-supervised embeddings were used as the input features of the SE model, and the performance improvement was somewhat limited. This may be because most of the self-supervised models were trained by maximizing the prediction probability of the target class. Features aimed for classification tasks may not fully retain detailed speech information and thus the suitability for direct application on SE generation tasks can be questioned.

To solve the aforementioned problem, we concatenated the $\log 1p$ spectrogram with SSL embedding to preserve useful information in speech. For the spectrogram, the window size and hop length are set as 400 and 160, respectively. As SSL features have twice the stride length of the spectrogram, we duplicated the latent representation to align the lengths of the embedding and spectrogram. To show the impact of adding the spectrogram as extra model input, comparisons were made in Table~\ref{table: SSL_models}. As the baselines, we trained the SE model with the embeddings from 1) the last hidden layer (LL) and 2) the weighted sum (WS) of all the embedding layers with the learned weights. 

In Table~\ref{table: SSL_models}, as a conventional method, we first show the results of applying only $\log 1p$ spectrogram as the model input. For SSL embedding, we prepared three different models: wav2vec 2.0, HuBERT, and WavLM-Base. From the table, we can first observe that using the last hidden layer of SSL models did not bring benefits compared to the spectrogram features. However, using WS of all the embedding layers can significantly improve the performance, implying that other layers in the SSL models also contain useful information for SE. For both the LL and WS, when the $\log 1p$ spectrogram is concatenated with the SSL features, the performance can improve further. The best performance results from the cross-domain feature (WS + $\log 1p$). Thus, this verified that including acoustic features improved the SE performance.

\begin{table}[th]
  \caption{Scores for various combinations of $\log 1p$ and latent representations under different SSL models. LL denotes the last layer embedding and WS as the weighted sum of all SSL layers.}
  \label{table: SSL_models}
  \centering
  \begin{tabular}{l@{}ccccc}
    \toprule
    \multicolumn{1}{c}{\textbf{Method}} & \multicolumn{1}{c}{\textbf{PESQ}} & \multicolumn{1}{c}{\textbf{CSIG}} & \multicolumn{1}{c}{\textbf{CBAK}} & \multicolumn{1}{c}{\textbf{COVL}} & \multicolumn{1}{c}{\textbf{STOI}} \\
    \midrule
    Noisy       &1.97   &3.35  & 2.44  &2.63  &0.915 \\
    $\log 1p$   &2.75   &4.15  & 3.36  &3.46  &0.944 \\

    \cmidrule(r){1-6}
    \multicolumn{6}{c}{\textbf{wav2vec 2.0- Base}}\\
    \cmidrule(r){1-6}
    LL  & 2.71   & 4.10  & 3.26  & 3.40  & 0.942 \\
    LL + $\log 1p$         & 2.91   & 4.29  & 3.42  & 3.60  & 0.948 \\
    WS    & 2.85   & 4.22  & 3.38  & 3.54  & 0.946 \\
    WS + $\log 1p$            & \textbf{2.94}   & \textbf{4.32}  & \textbf{3.45}  & \textbf{3.64}  & \textbf{0.949} \\

        \cmidrule(r){1-6}
    \multicolumn{6}{c}{\textbf{HuBERT- Base}}\\
    \cmidrule(r){1-6}
    LL   & 2.67   & 4.04  & 3.20  & 3.35  & 0.942 \\
    LL + $\log 1p$         & 2.94   & 4.31  & 3.46  & 3.63  & 0.948 \\
    WS   & 2.84   & 4.23  & 3.37  & 3.54  & 0.947 \\
    WS + $\log 1p$             & \textbf{2.98}   & \textbf{4.34}  & \textbf{3.48}  & \textbf{3.67}  & \textbf{0.949} \\

    \cmidrule(r){1-6}
    \multicolumn{6}{c}{\textbf{WavLM- Base}}\\
    \cmidrule(r){1-6}
    LL   & 2.74   & 4.05  & 3.22  & 3.39  & 0.944 \\
    LL + $\log 1p$         & 2.94   & 4.32  & 3.44  & 3.64  & 0.950 \\
    WS   & 2.90   & 4.28  & 3.43  & 3.59  & 0.949 \\
    WS + $\log 1p$          & \textbf{3.05}   & \textbf{4.40}  & \textbf{3.52}  & \textbf{3.74}  & \textbf{0.952} \\


    \bottomrule
  \end{tabular}
\end{table}

\subsubsection{Noise robustness and learned weights of weighted sum representation}\label{subsec: Noise robustness}
\begin{figure}[ht]
  \centering
  \includegraphics[width=\linewidth]{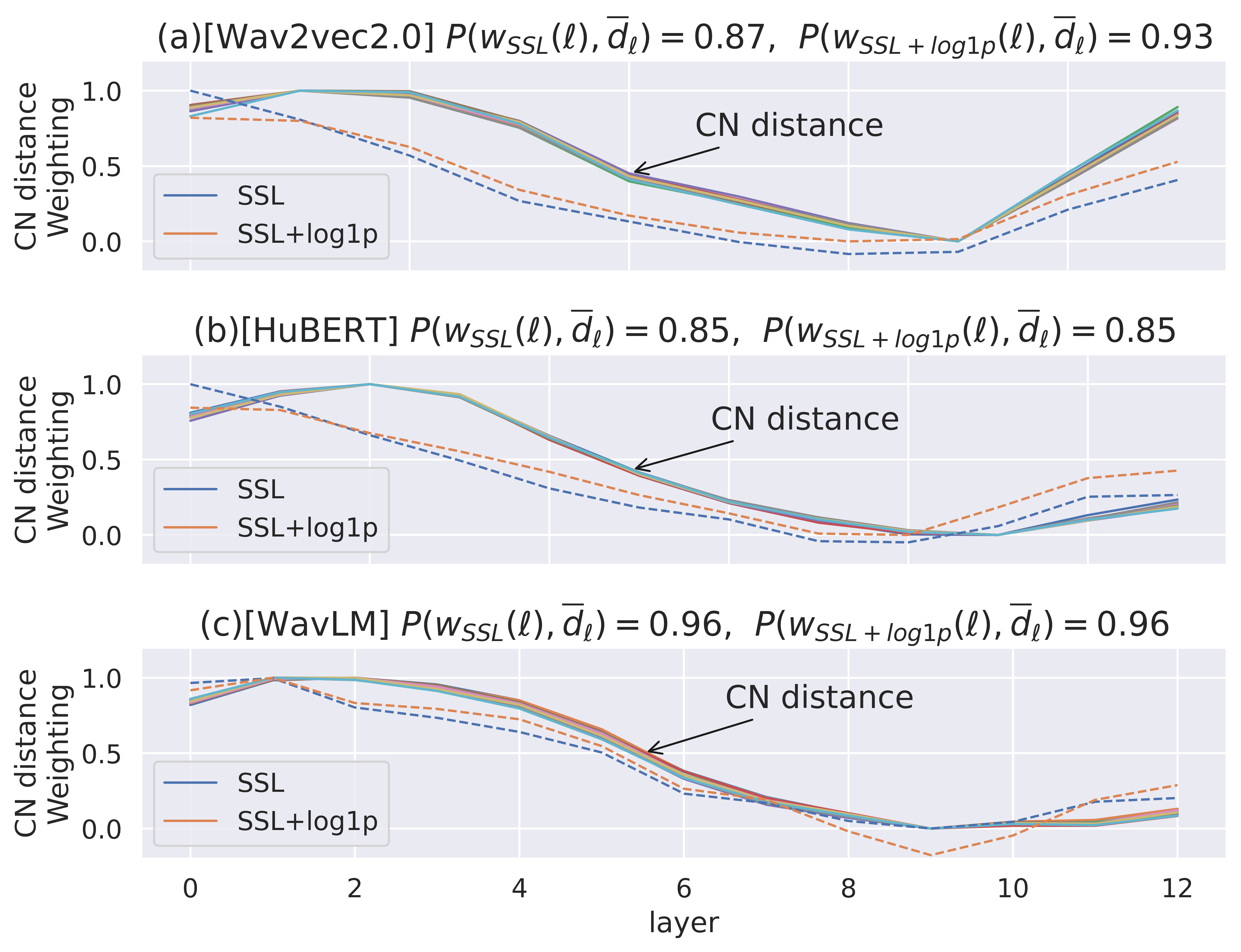}
  \caption{Correlation between the CN distance (solid line bundle, see Eq.~(\ref{E:avg CN distance})) and the learned weights $w(\ell)$ (blue/orange dotted line) of Eq.~(\ref{E:weightedsum}) using SSL only and SSL $+ \log 1p$ features, respectively.}
  \label{fig:CN_dist}
\end{figure}

Fig.~\ref{fig:CN_dist} shows the CN distances (solid lines) and layer weightings (dotted lines) for the various SSL models. Each solid line
follows from Eq.~(\ref{E:distance}) with randomly sampled inputs $(\mathbf{X}_c, \mathbf{X}_n)$ out of the training set, as described in Sec.~\ref{subsec:Noise robustness analysis}. The averaging CN distance curve is then calculated by Eq.~(\ref{E:avg CN distance}) to compare with the trend of layer weighting curves. To ensure the distance $d_{f, \ell}$ falling into $[0, 1]$, the min-max normalization was employed after the computation of Eq.~(\ref{E:distance}).

An SSL model $f$ results in layer weighting curves (Fig. ~\ref{fig:CN_dist}), $\ell \mapsto w_{\text{SSL}}^{(f)}(\ell)$ and $ \ell \mapsto w_{\text{SSL+log1p}}^{(f)}(\ell)$, corresponding to the use of the SSL feature and the cross-domain features, respectively.
As \cite{9688093} reported that there existed some peculiarity in the last two layers of wav2vec 2.0, we removed the last two layers in wav2vec 2.0 for comparison here. In these three models, it was observed that the layer weightings and CN distance $\overline{d}_{f, \ell}$ are highly correlated ($\geq 0.85$) via the Pearson correlation $P(w_{\text{SSL}}^{(f)}(\ell), \overline{d}_{f, \ell})$, for $f \in \{ \text{WavLM}, \text{HuBERT}, \text{wav2vec 2.0} \}$. Thus, we concluded that large CN distances might provide more information.

\subsubsection{Fine-tuning SSL models}
From Table~\ref{table: SSL_models}, it can be observed that applying WavLM as an SSL model outperforms the other two methods. Under the same model size, WavLM achieves 3.05 and 0.952 in PESQ and STOI, respectively, higher than HuBERT (PESQ: 2.99/ STOI: 0.949) and Wav2vec 2.0 (PESQ: 2.95/ STOI: 0.949). Henceforth, we shall only adopt WavLM for the discussion of the fine-tuning experiments later.

Table~\ref{table:2} presents the fine-tuned results of WavLM, for both WavLM-Base and WavLM-Large, the best performance comes from PF with $\log 1p$. It can also be observed that EF with $\log 1p$ did not improve much compared with PF. Fine-tune models with $\log 1p$ have only incremental improvements compared with the original pre-trained models (cf. Table~\ref{table: SSL_models}). From these observations, we believe that the SSL models after fine-tune acquired better latent representations for SE, such that the additional $\log 1p$ features can only provide limited extra information for enhancement. We also observed that the performance  sharply degraded if we used a model with the same architecture but trained from scratch (TFS), which indicated that pre-training certainly contributes. Some SOTA SE models using a pre-trained SSL model are listed in Table~\ref{table:2} for comparison. 

\begin{figure}[t]
  \centering
  \includegraphics[width=\linewidth]{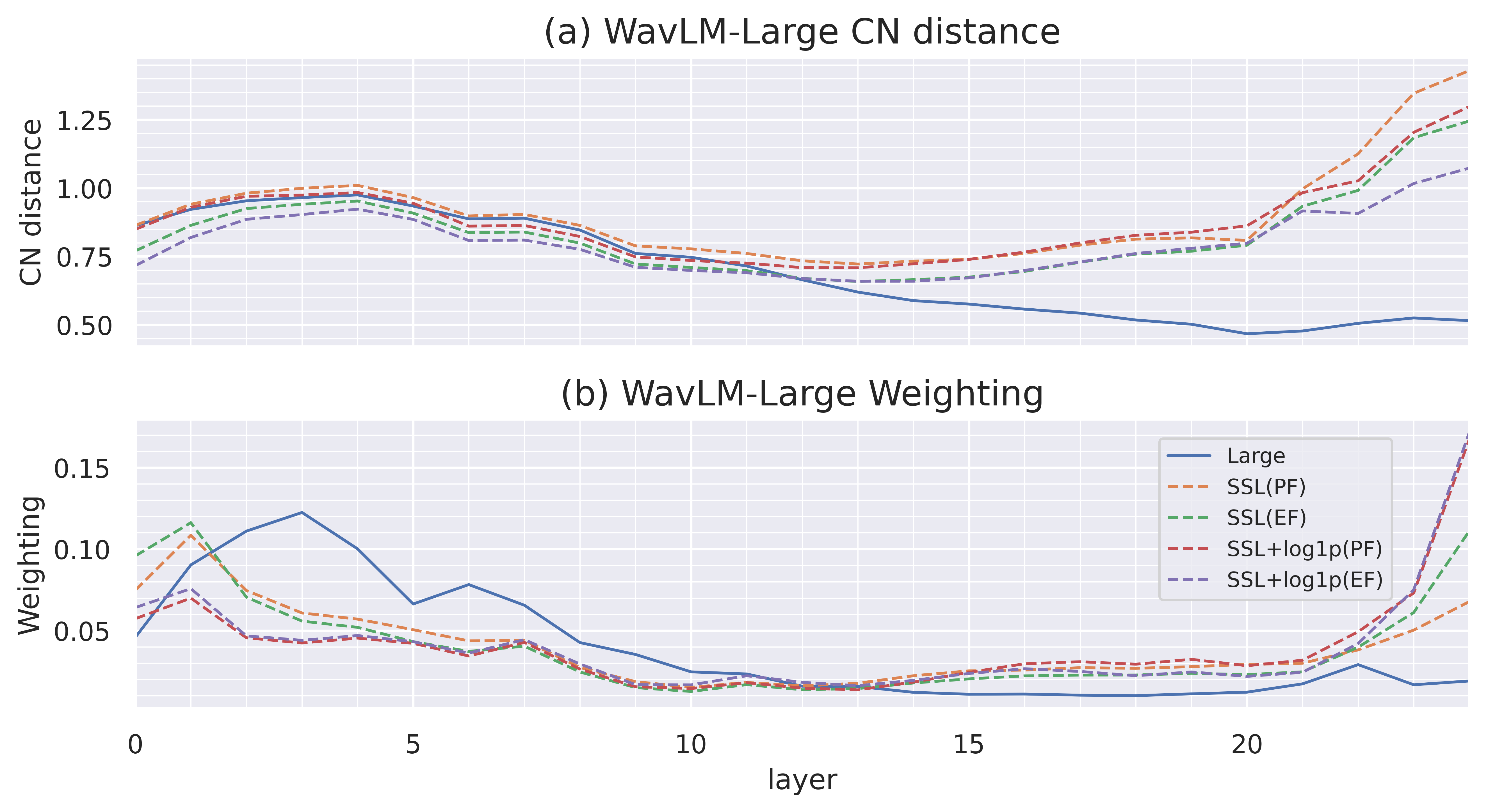}
  \caption{The averaging CN distance (a) and the layer weighting curves (b) before (solid line) and after (dotted line) different fine-tuning strategies. Plot (a) and (b) shares the same legend.}
  \label{fig:finetune}
\end{figure}

In addition to evaluating the performance of SSL fine-tuned model, the corresponding CN distances were also calculated, as shown in Fig.~\ref{fig:finetune}(a). The figure reveals that the CN distances in the first and last few layers increased after fine-tuning. This was particularly obvious in the last few layers. From Fig.~\ref{fig:finetune}(b), we saw similar trends for learned weights, which verify the observation that large CN distances may provide more information as given in Sec.~\ref{subsec: Noise robustness}. 

Another interesting finding in Fig.~\ref{fig:finetune}(b) is that when using the SSL representations only (orange and green dotted lines), the weights in the first few layers also increase and become larger than that of SSL+$\log 1p$ (red and purple dotted lines). We argue that after fine-tuning without $\log 1p$ input, the first few layers learn more information about raw acoustic features. Since the first few layers can keep more local information now, the performance gained by $\log 1p$ feature becomes much smaller, as shown in Table~\ref{table:2}.

\begin{table}[th]
  \caption{Scores for different SSL fine-tuning strategies. TFS denotes training from scratch (random initial weights in WavLM).}
  \label{table:2}
  \centering
  \begin{tabular}{l@{}ccccc}
    \toprule
    \multicolumn{1}{c}{\textbf{Method}} & \multicolumn{1}{c}{\textbf{PESQ}} & \multicolumn{1}{c}{\textbf{CSIG}} & \multicolumn{1}{c}{\textbf{CBAK}} & \multicolumn{1}{c}{\textbf{COVL}} & \multicolumn{1}{c}{\textbf{STOI}} \\
    \midrule
    SSPF~\cite{qiu2021self}                  & 3.13   & 4.30   & 3.61  & 3.72    & 0.950 \\
    PERL~\cite{kataria2021perceptual}  & 3.04   & 4.23   & 3.42   & 3.63   & 0.947 \\
    PFPL~\cite{hsieh2020improving}           & 3.15   & 4.18   & 3.60   & 3.67   & 0.950 \\
    Huang~\cite{ZiliSSL}                          & 2.80   & N/A      & N/A      & N/A      & 0.945 \\
    \midrule
    \multicolumn{6}{c}{\textbf{WavLM- Base (WS)}}\\
    \cmidrule(r){1-6}
    TFS     & 2.83   &4.21  & 3.41   &3.53  & 0.946 \\
    PF  & 3.09   & 4.42   & 3.54   & 3.77   & 0.955 \\
    EF  & 3.11   & 4.44   & 3.56   & 3.79   & 0.955 \\
    $\log 1p$    & 3.05   & 4.40   & 3.52  & 3.74    & 0.952 \\
    $\log 1p$ + PF   & \textbf{3.16}  & \textbf{4.50}  & \textbf{3.57}   & \textbf{3.86}   & \textbf{0.956} \\
    $\log 1p$ + EF   & 3.12  & 4.49  & 3.56   & 3.83   & 0.956 \\
    

    \cmidrule(r){1-6}
    \multicolumn{6}{c}{\textbf{WavLM- Large (WS)}}\\
    \cmidrule(r){1-6}
    TFS     & 2.87   &4.24  & 3.41   &3.57  & 0.945 \\
    PF  & 3.14 & 4.47 & 3.57 & 3.82 & 0.957 \\
    EF  & 3.17 & 4.49 & 3.58 & 3.85 & 0.956 \\
    $\log 1p$  & 3.09 & 4.45 & 3.53 & 3.79 & 0.954 \\
    $\log 1p$ + PF   & \textbf{3.20} & \textbf{4.53} & \textbf{3.60} & \textbf{3.88} & \textbf{0.957} \\
    $\log 1p$ + EF  & 3.15 & 4.50 & 3.59 & 3.85 & 0.957 \\

    \bottomrule
  \end{tabular}
\end{table}

\section{Conclusions}

In this study, we propose two techniques: 1) utilizing cross-domain features as model inputs, 2) fine-tuning the SSL model with the SE task to compensate for the information loss in the SSL features. The results met our expectations in that the SE performance with the cross-domain feature was significantly improved than using only SSL representation or the spectrogram feature. Compared to SOTA SSL-based SE methods, our proposal for fine-tuning an SSL representation with the SE task can outperform them in PESQ, CSIG, and COVL without invoking complicated network architectures or training flow.  Furthermore, we studied the relationship between the noise robustness of SSL representation and the importance of SE. Our observation showed that less noise-robust SSL features possess higher corresponding importance. Although this fact appeared counter-intuitive, we addressed the rationale behind the scenes. In the end, we found that the SSL representation with the weighted-sum method has proven superior to the spectrogram feature. However, training an SSL representation to entirely replace raw acoustic features is yet to be explored. Our findings serve to help train SE-related SSL representation in the future.

\bibliographystyle{IEEEtran}
\bibliography{refs}

\begin{thebibliography}{10}
\providecommand{\url}[1]{#1}
\csname url@samestyle\endcsname
\providecommand{\newblock}{\relax}
\providecommand{\bibinfo}[2]{#2}
\providecommand{\BIBentrySTDinterwordspacing}{\spaceskip=0pt\relax}
\providecommand{\BIBentryALTinterwordstretchfactor}{4}
\providecommand{\BIBentryALTinterwordspacing}{\spaceskip=\fontdimen2\font plus
\BIBentryALTinterwordstretchfactor\fontdimen3\font minus
  \fontdimen4\font\relax}
\providecommand{\BIBforeignlanguage}[2]{{%
\expandafter\ifx\csname l@#1\endcsname\relax
\typeout{** WARNING: IEEEtran.bst: No hyphenation pattern has been}%
\typeout{** loaded for the language `#1'. Using the pattern for}%
\typeout{** the default language instead.}%
\else
\language=\csname l@#1\endcsname
\fi
#2}}
\providecommand{\BIBdecl}{\relax}
\BIBdecl

\bibitem{xu2014regression}
Y.~Xu, J.~Du, L.-R. Dai, and C.-H. Lee, ``A regression approach to speech
  enhancement based on deep neural networks,'' \emph{IEEE/ACM Transactions on
  Audio, Speech, and Language Processing}, vol.~23, no.~1, pp. 7--19, 2014.

\bibitem{lu2013speech}
X.~Lu, Y.~Tsao, S.~Matsuda, and C.~Hori, ``Speech enhancement based on deep
  denoising autoencoder.'' in \emph{Proc. Interspeech 2013}.

\bibitem{wang2018supervised}
D.~Wang and J.~Chen, ``Supervised speech separation based on deep learning: An
  overview,'' \emph{IEEE/ACM Transactions on Audio, Speech, and Language
  Processing}, vol.~26, no.~10, pp. 1702--1726, 2018.

\bibitem{wang2021speech}
Y.~Wang, J.~Han, T.~Zhang, and D.~Qing, ``Speech enhancement from fused
  features based on deep neural network and gated recurrent unit network,''
  \emph{EURASIP Journal on Advances in Signal Processing}, vol. 2021, no.~1,
  pp. 1--19, 2021.

\bibitem{valin2020perceptually}
J.-M. Valin, U.~Isik, N.~Phansalkar, R.~Giri, K.~Helwani, and A.~Krishnaswamy,
  ``A perceptually-motivated approach for low-complexity, real-time enhancement
  of fullband speech,'' \emph{arXiv preprint arXiv:2008.04259}, 2020.

\bibitem{choi2018phase}
H.-S. Choi, J.-H. Kim, J.~Huh, A.~Kim, J.-W. Ha, and K.~Lee, ``Phase-aware
  speech enhancement with deep complex u-net,'' in \emph{International
  Conference on Learning Representations}, 2018.

\bibitem{qi2020tensor}
J.~Qi, H.~Hu, Y.~Wang, C.-H.~H. Yang, S.~M. Siniscalchi, and C.-H. Lee,
  ``Tensor-to-vector regression for multi-channel speech enhancement based on
  tensor-train network,'' in \emph{Proc. ICASSP 2020}.

\bibitem{kumar2016speech}
A.~Kumar and D.~Florencio, ``Speech enhancement in multiple-noise conditions
  using deep neural networks,'' \emph{arXiv preprint arXiv:1605.02427}, 2016.

\bibitem{sivaraman2021personalized}
A.~Sivaraman, S.~Kim, and M.~Kim, ``Personalized speech enhancement through
  self-supervised data augmentation and purification,'' \emph{arXiv preprint
  arXiv:2104.02018}, 2021.

\bibitem{kim2021specmix}
G.~Kim, D.~K. Han, and H.~Ko, ``Specmix: A mixed sample data augmentation
  method for training withtime-frequency domain features,'' \emph{arXiv
  preprint arXiv:2108.03020}, 2021.

\bibitem{liu2022audio}
S.~Liu, A.~Mallol-Ragolta, E.~Parada-Cabeleiro, K.~Qian, X.~Jing, A.~Kathan,
  B.~Hu, and B.~W. Schuller, ``Audio self-supervised learning: A survey,''
  \emph{arXiv preprint arXiv:2203.01205}, 2022.

\bibitem{oord2018representation}
A.~v.~d. Oord, Y.~Li, and O.~Vinyals, ``Representation learning with
  contrastive predictive coding,'' \emph{arXiv preprint arXiv:1807.03748},
  2018.

\bibitem{schneider2019wav2vec}
S.~Schneider, A.~Baevski, R.~Collobert, and M.~Auli, ``wav2vec: Unsupervised
  pre-training for speech recognition,'' \emph{arXiv preprint
  arXiv:1904.05862}, 2019.

\bibitem{hsu2021hubert}
W.-N. Hsu, B.~Bolte, Y.-H.~H. Tsai, K.~Lakhotia, R.~Salakhutdinov, and
  A.~Mohamed, ``Hubert: Self-supervised speech representation learning by
  masked prediction of hidden units,'' \emph{IEEE/ACM Transactions on Audio,
  Speech, and Language Processing}, vol.~29, pp. 3451--3460, 2021.

\bibitem{wang2021fine}
Y.~Wang, A.~Boumadane, and A.~Heba, ``A fine-tuned wav2vec 2.0/hubert benchmark
  for speech emotion recognition, speaker verification and spoken language
  understanding,'' \emph{arXiv preprint arXiv:2111.02735}, 2021.

\bibitem{cooper2022generalization}
E.~Cooper, W.-C. Huang, T.~Toda, and J.~Yamagishi, ``Generalization ability of
  mos prediction networks,'' in \emph{Proc. ICASSP 2022}.

\bibitem{nguyen2020investigating}
H.~Nguyen, F.~Bougares, N.~Tomashenko, Y.~Est{\`e}ve, and L.~Besacier,
  ``Investigating self-supervised pre-training for end-to-end speech
  translation,'' in \emph{Proc. Interspeech 2020}.

\bibitem{ZiliSSL}
Z.~Huang, S.~Watanabe, S.-w. Yang, P.~Garc{\'\i}a, and S.~Khudanpur,
  ``Investigating self-supervised learning for speech enhancement and
  separation,'' in \emph{Proc. ICASSP 2022}.

\bibitem{chung2019unsupervised}
Y.-A. Chung, W.-N. Hsu, H.~Tang, and J.~Glass, ``An unsupervised autoregressive
  model for speech representation learning,'' in \emph{Proc. Interspeech 2019}.

\bibitem{liu2020mockingjay}
A.~T. Liu, S.-w. Yang, P.-H. Chi, P.-c. Hsu, and H.-y. Lee, ``Mockingjay:
  Unsupervised speech representation learning with deep bidirectional
  transformer encoders,'' in \emph{Proc. ICASSP 2020}.

\bibitem{ling2020decoar}
S.~Ling and Y.~Liu, ``Decoar 2.0: Deep contextualized acoustic representations
  with vector quantization,'' \emph{arXiv preprint arXiv:2012.06659}, 2020.

\bibitem{tagliasacchi2020pre}
M.~Tagliasacchi, B.~Gfeller, F.~de~Chaumont~Quitry, and D.~Roblek,
  ``Pre-training audio representations with self-supervision,'' \emph{IEEE
  Signal Processing Letters}, vol.~27, pp. 600--604, 2020.

\bibitem{chen2021wavlm}
S.~Chen, C.~Wang, Z.~Chen, Y.~Wu, S.~Liu, Z.~Chen, J.~Li, N.~Kanda,
  T.~Yoshioka, X.~Xiao \emph{et~al.}, ``Wav{LM}: Large-scale self-supervised
  pre-training for full stack speech processing,'' \emph{arXiv preprint
  arXiv:2110.13900}, 2021.

\bibitem{huang2022spiral}
W.~Huang, Z.~Zhang, Y.~T. Yeung, X.~Jiang, and Q.~Liu, ``Spiral:
  Self-supervised perturbation-invariant representation learning for speech
  pre-training,'' \emph{arXiv preprint arXiv:2201.10207}, 2022.

\bibitem{ravanelli2020multi}
M.~Ravanelli, J.~Zhong, S.~Pascual, P.~Swietojanski, J.~Monteiro, J.~Trmal, and
  Y.~Bengio, ``Multi-task self-supervised learning for robust speech
  recognition,'' in \emph{Proc. ICASSP 2020}.

\bibitem{yang2021superb}
S.-w. Yang, P.-H. Chi, Y.-S. Chuang, C.-I.~J. Lai, K.~Lakhotia, Y.~Y. Lin,
  A.~T. Liu, J.~Shi, X.~Chang, G.-T. Lin \emph{et~al.}, ``Superb: Speech
  processing universal performance benchmark,'' \emph{arXiv preprint
  arXiv:2105.01051}, 2021.

\bibitem{zezario2021deep}
R.~E. Zezario, S.-W. Fu, F.~Chen, C.-S. Fuh, H.-M. Wang, and Y.~Tsao, ``Deep
  learning-based non-intrusive multi-objective speech assessment model with
  cross-domain features,'' \emph{arXiv preprint arXiv:2111.02363}, 2021.

\bibitem{wang2020self}
Y.-C. Wang, S.~Venkataramani, and P.~Smaragdis, ``Self-supervised learning for
  speech enhancement,'' \emph{arXiv preprint arXiv:2006.10388}, 2020.

\bibitem{9053925}
R.~E. Zezario, T.~Hussain, X.~Lu, H.-M. Wang, and Y.~Tsao, ``Self-supervised
  denoising autoencoder with linear regression decoder for speech
  enhancement,'' in \emph{Proc. ICASSP 2020}.

\bibitem{hsieh2020improving}
T.-A. Hsieh, C.~Yu, S.-W. Fu, X.~Lu, and Y.~Tsao, ``Improving perceptual
  quality by phone-fortified perceptual loss using wasserstein distance for
  speech enhancement,'' \emph{arXiv preprint arXiv:2010.15174}, 2020.

\bibitem{kataria2021perceptual}
S.~Kataria, J.~Villalba, and N.~Dehak, ``Perceptual loss based speech denoising
  with an ensemble of audio pattern recognition and self-supervised models,''
  in \emph{Proc. ICASSP 2021}.

\bibitem{sun2021boosting}
T.~Sun, S.~Gong, Z.~Wang, C.~D. Smith, X.~Wang, L.~Xu, and J.~Liu, ``Boosting
  the intelligibility of waveform speech enhancement networks through
  self-supervised representations,'' in \emph{Proc. ICMLA 2021}.

\bibitem{tsai2022superb}
H.-S. Tsai, H.-J. Chang, W.-C. Huang, Z.~Huang, K.~Lakhotia, S.-w. Yang,
  S.~Dong, A.~T. Liu, C.-I.~J. Lai, J.~Shi \emph{et~al.}, ``Superb-sg: Enhanced
  speech processing universal performance benchmark for semantic and generative
  capabilities,'' \emph{arXiv preprint arXiv:2203.06849}, 2022.

\bibitem{qiu2021self}
Y.~Qiu, R.~Wang, S.~Singh, Z.~Ma, and F.~Hou, ``Self-supervised learning based
  phone-fortified speech enhancement,'' \emph{Proc. Interspeech 2021}.

\bibitem{9688093}
A.~Pasad, J.-C. Chou, and K.~Livescu, ``Layer-wise analysis of a
  self-supervised speech representation model,'' in \emph{Proc. ASRU 2021}.

\bibitem{fu2020boosting}
S.-W. Fu, C.-F. Liao, T.-A. Hsieh, K.-H. Hung, S.-S. Wang, C.~Yu, H.-C. Kuo,
  R.~E. Zezario, Y.-J. Li, S.-Y. Chuang \emph{et~al.}, ``Boosting objective
  scores of a speech enhancement model by metricgan post-processing,'' in
  \emph{Proc. APSIPA 2020}.

\bibitem{valentini2016investigating}
C.~Valentini-Botinhao, X.~Wang, S.~Takaki, and J.~Yamagishi, ``Investigating
  rnn-based speech enhancement methods for noise-robust text-to-speech.'' in
  \emph{Proc. SSW 2016}.

\bibitem{liu2019supervised}
Y.~Liu, H.~Zhang, X.~Zhang, and L.~Yang, ``Supervised speech enhancement with
  real spectrum approximation,'' in \emph{Proc. ICASSP 2019}.

\end{thebibliography}

\end{document}